\documentclass[useAMS,usenatbib]{mn2e}
\usepackage{graphicx,amsmath,rotating,fleqn,color}

\newcommand{\figdir}
  {}
\newlength{\figwidth}
\newlength{\thinfigwidth}
\newcommand{\unit}[1]
  {{\mbox{\rm\,\,#1}}}
\newcommand{\percent}
  {\,\%}

\newcommand{\skyhat}
  {\hat{\vect{r}}}
\newcommand{\prob}
  {{\rmn{Pr}}}
\newcommand{\gzk}
  {{\rm{GZK}}}
\newcommand{\PAO}
  {PAO}
\newcommand{\VCV}
  {VCV}

\newcommand{\obs}
  {{\rm{obs}}}
\newcommand{\uhemin}
  {{\rm{min}}}
\newcommand{\arr}
  {{\rm{arr}}}
\newcommand{\emit}
  {{\rm{emit}}}
\newcommand{\rate}
  {\Gamma}
\newcommand{\ratebkg}
  {R_{\rmn{bkg}}}
\newcommand{\ratesrc}
  {\Gamma_{\rmn{src}}}
\newcommand{\px}
  {p}
\newcommand{\psrc}
  {P_{\rmn{src}}}
\newcommand{\npx}
  {N_{\rmn{p}}}
\newcommand{\data}
  {{\rmn{data}}}
\newcommand{\ncrpx}
  {N_{{\rmn{c}},p}}
\newcommand{\ncrpxexpsrc}
  {\overline{N}_{{\rmn{src}},p}}
\newcommand{\ncrpxexpbkg}
  {\overline{N}_{{\rmn{bkg}},p}}
\newcommand{\databins}
  {\{\ncrpx\}}
\newcommand{\nsrc}
  {N_{\rmn{s}}}
\newcommand{\src}
  {s}
\newcommand{\source}
  {{\rmn{src}}}

\newcommand{\fagn}
  {F_{\rmn{AGN}}}

\newcommand{\redshift}
  {z}

\newcommand{\cray}
  {c}
\newcommand{\cosmicray}
  {CR}
\newcommand{\cosmicrays}
  {CRs}
\newcommand{\ncray}
  {N_{\rmn{c}}}
\newcommand{\uhecr}
  {UHECR}
\newcommand{\uhecrs}
  {UHECRs}
\newcommand{\cry}
  {CR}
\newcommand{\crs}
  {CRs}

\newcommand{\agn}
  {AGN}
\newcommand{\agns}
  {AGNs}
\newcommand{\exposure}
  {\epsilon}

\newcommand{\etal}
  {et al.}
\newcommand{\eg}
  {e.g.}
\newcommand{\cf}
  {cf.}
\newcommand{\ie}
  {i.e.}
\newcommand{\diff}
  {{\rmn{d}}}
\newcommand{\vect}[1]
  {\mbox{\boldmath ${#1}$}}

\newcommand{\eq}[1]
  {Eq.~(\ref{eq:#1})}
\newcommand{\eqs}[1]
  {Eqs~(\ref{eq:#1})}

\newcommand{\sect}[1]
  {Section~\ref{section:#1}}

\newcommand{\tabl}[1]
  {{\mbox Table~\ref{table:#1}}}

\newcommand{\fig}[1]
  {Fig.~\ref{figure:#1}}
\newcommand{\figs}[1]
  {Figs.~\ref{figure:#1}}

\title
  [The highest energy cosmic rays]
  {A Bayesian analysis of the 27 highest energy cosmic rays detected 
    by the Pierre Auger Observatory}
\author
  [L.\ J. Watson \etal]
  {Laura J.\ Watson\thanks{E-mail: l.watson09@imperial.ac.uk},
  Daniel J.\ Mortlock
  and Andrew H.\ Jaffe
 \vspace*{2mm}\\
Astrophysics Group, Imperial College London, Blackett Laboratory,
  Prince Consort Road, London SW7 2AZ, U.K.\\
  }
\begin{document}

\date{Received 2010 October 5}

\pagerange{\pageref{firstpage}--\pageref{lastpage}} \pubyear{2011}

\maketitle

\label{firstpage}

\begin{abstract}
It is possible that ultra-high energy cosmic rays (\uhecrs) are generated by active galactic nuclei (\agns), but there is currently no conclusive evidence for this hypothesis.  Several reports of correlations between the arrival directions of \uhecrs\ and the positions of nearby \agns\ have been made, the strongest detection coming from a sample of 27 \uhecrs\ detected by the Pierre Auger Observatory (\PAO). However, the PAO results were based on a statistical methodology that not only ignored some relevant information (most obviously the \uhecr\ arrival energies but also some of the information in the arrival directions) but also involved some problematic fine-tuning of the correlation parameters.  Here we present a fully Bayesian analysis of the \PAO\ data (collected before 2007 September), which makes use of more of the available information, and find that a fraction $\fagn = 0.15 ^{+ 0.10}_{- 0.07}$ of the \uhecrs\ originate from known \agns\ in the Veron-Cetty \& Veron (\VCV) catalogue.  The hypothesis that all the \uhecrs\ come from \VCV\ \agns\ is ruled out, although there remains a small possibility that the \PAO-\agn\ correlation is coincidental ($\fagn = 0.15$ is 200 times as probable as $\fagn = 0.00$).
\end{abstract}

\begin{keywords}
cosmic rays -- methods: statistics -- galaxies: active -- surveys 
\end{keywords}


\section{Introduction}
\label{section:intro}

Cosmic rays (\cosmicrays) are highly accelerated protons and nuclei that reach Earth with arrival energies in the wide range $10^8 \unit{eV} \la E_\arr \la 10^{20} \unit{eV}$ (see, \eg\ \citealt{Stoker:2009}). However the origin of ultra-high energy cosmic rays (UHECRs) with $E_\arr \ga 10^{19} \unit{eV}$, in particular, remains uncertain. The most promising theory is that \uhecrs\ are generated by active galactic nuclei (AGNs), and there are several physical models to motivate this idea (\eg\ \citealt{Protheroe_Szabo:1992,Diehl:2009}), but this hypothesis requires empirical verification.

A number of difficulties hinder efforts to gain experimental evidence about \uhecrs. The most fundamental problem is that \cosmicrays\ are deflected by the Galaxy's magnetic field: the arrival directions of lower energy protons are essentially independent of their point of origin, although \uhecrs\ are expected to be deflected by no more than a few degrees (\eg\ \citealt{Dolag_etal:2005,MedinaTanco_etal:1998}). It is also problematic that \uhecrs\ are very rare, with the observed number flux falling off with energy as $\diff \rate_\obs / \diff E_\arr \simeq [E_\arr / (10^{19} \unit{eV})]^{-2.6} \unit{s}^{-1} \unit{m}^{-2} \unit{sr}^{-1}$ (\eg\ \citealt{AugerCollab:2010}).  The fall-off is expected to be even more extreme above $E_{\gzk} \simeq 5 \times 10^{19} \unit{eV}$, as protons at these energies interact with cosmic microwave background (CMB) photons to produce pions \citep[hereafter \gzk]{Greisen:1966,Zatsepin_Kuzmin:1966}. The \gzk\ mean free path between interactions for an $E \simeq 10^{20} \unit{eV}$ proton is only about $4 \unit{Mpc}$, and each interaction typically reduces a \cosmicray's energy by approximately $20 \unit{per cent}$ (\eg\ \citealt{Rachen_Biermann:1993}), so any observed \uhecrs\ must have originated within an effective `\gzk\ horizon' of about $100 \unit{Mpc}$.  (If, alternatively, \uhecrs\ are primarily Fe nuclei the \gzk\ horizon is expected to be even smaller, although the deflection due to magnetic fields is greatly increased.)  The \gzk\ effect reduces the number of detectable \uhecrs, but a fortunate consequence is that it also reduces the number of plausible \agn\ sources to the few thousand with distances of $D \la 100 \unit{Mpc}$ or, equivalently, redshifts of $z \la 0.03$.  This makes it plausible to search for a correlation between the arrival directions of \uhecrs\ and locations of local \agns, provided sufficiently many \uhecrs\ can be observed.

The problem of the low \uhecr\ arrival rate can only be overcome by using a large collecting area, and by observing for long periods of time. At present, the largest \cosmicray\ observatory is the Pierre Auger Observatory (\PAO; \citealt{AugerCollab:2004}), which has been operational since 2004 January. During its first $3.6$ years of observing, the \PAO\ made reliable detections of the arrival directions and energies of 81 \uhecrs, of which 27 had an (esimated) arrival energy of $E_\obs \geq 5.7 \times 10^{19} \unit{eV}$ \citep{AugerCollab:2007b}. These 27 events were found to be strongly correlated with a sample of local \agns\ in the \citet[hereafter \VCV]{VeronCetty:2006} catalogue; this was the first strong empirical support for the hypothesis that \uhecrs\ are generated by \agns. The \PAO\ has continued to operate in the time since these results were obtained; the latest data \citep{AugerCollab:2010b} show a much weaker correlation. 
As discussed in \cite{Beatty_Westerhoff:2009}, there have been many attempts to find a correlation between \agns\ and \uhecrs, using a variety of techniques and data, such as those reported by \cite{Nemmen_etal:2010}, \cite{AugerCollab:2008,AugerCollab:2007b}, \cite{HiresCollab:2008}, \cite{Ghisellini_etal:2008} and \cite{George_etal:2008}. In particular, \cite{HiresCollab:2008} claim no significant correlation.

Given the small numbers of \uhecrs\ on which the \PAO\ results are based, some care must be taken with statistical methods,
both to ensure that all the available information is utilised and to avoid over-interpretation. These aims can be achieved by adopting a Bayesian approach in which the relevant stochastic processes (\eg\ the \gzk\ interactions of the \uhecrs\ with the CMB, deflection by the Galaxy's magnetic field, measurement errors) are explicitly modelled. Even though the details of some of these processes are not well known (most relevantly the strength of the magnetic fields and the energy calibration of the \uhecrs), such uncertainties can be accounted for by marginalisation. Whereas the simple correlation analysis of \cite{AugerCollab:2007b} ignores the arrival energy of the individual \uhecrs, implicitly assuming that the \gzk\ horizon is independent of $E_\arr$, a likelihood-based approach can incorporate the fact that the very highest energy events are expected to have come from the most nearby \agns.  Similarly, the use of circular angular regions to match \uhecr\ arrival directions and \agn\ positions is both sub-optimal (as real matches would tend to be more centrally concentrated) and potentially misleading (because any resultant statistic is as sensitive to physically implausible correlations as to the tighter angular matches that would be expected if the \agns\ were the \uhecrs' progenitors). 

While neither ignoring the individual \uhecrs' arrival energies nor using a hard effective \gzk\ horizon are necessarily inconsistent, both choices decrease the constraining power of the data-set. For instance, a strong prediction of the \agn\ hypothesis is not only that \uhecrs' arrival directions will be correlated with nearby \agns, but that every \agn-sourced \uhecr\ will be directly associated with at least one candidate source, and that the most energetic events will come from the closer \agns. Critically, it is possible to use a physical model of \uhecr\ generation, propagation and observation, hence extracting much more of the valuable information in a \uhecr\ data-set than is possible with other, more heuristic, analysis methods.

In this paper we take the first steps to developing a comprehensive Bayesian formalism for analysing \uhecr\ data.  Our starting point is to reanalyse the \uhecr\ and \agn\ samples used by \cite{AugerCollab:2007b}, changing only the statistical method.  Aside from providing an answer to the question of whether the 27 \PAO\ \uhecrs\ come from the local \VCV\ \agns, we will show directly how the results depend on the statistical method used to analyse such data-sets.  After describing the \uhecr\ and \agn\ samples in \sect{data}, our statistical method and \cry\ propagation model are presented in \sect{method}.  The results of applying this methodology are given in \sect{results} and the overall conclusions are summarised in \sect{conc}.


\section{Data}
\label{section:data}

The sample of \uhecrs\ (\sect{auger}) and the putative \agn\ sources (\sect{agn}) analysed here are the same as used by \cite{AugerCollab:2007b}.

\begin{figure*}
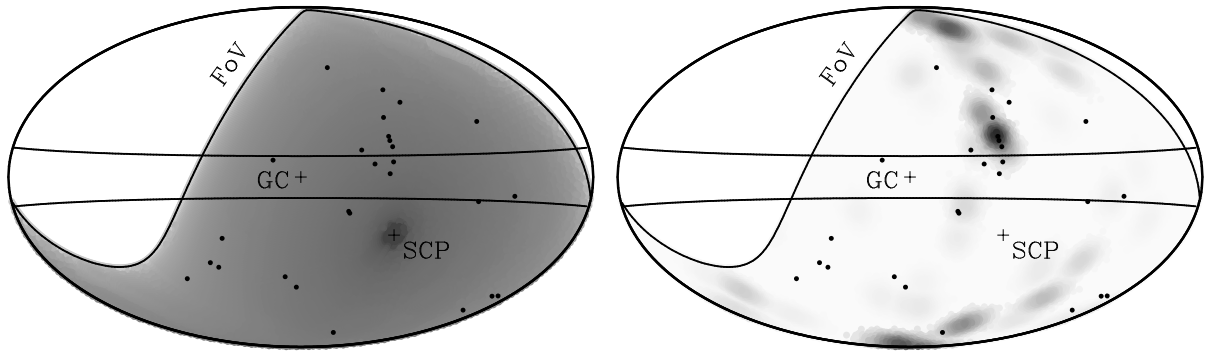

\includegraphics[width=\figwidth]{\figdir sky_bkg}
\includegraphics[width=\figwidth]{\figdir sky_agn}
\caption{The arrival directions of the $\ncray = 27$ \PAO\ \uhecrs\ (black points) and the source-weighted exposure (greyscale: darker indicates greater exposure) for the background-only model (left) and the \agn-only model (right), in Galactic coordinates.  The Galactic Centre (GC), South Celestial Pole (SCP) and \PAO's field of view (FoV) are all indicated.  Lines of constant Galactic latitude $|b| = 10\unit{deg}$ are also shown.}
\label{figure:skymap}
\end{figure*}

\subsection{\PAO\ observations of \uhecrs}
\label{section:auger}

The \PAO\ South is located near Malarg\"{u}e in Argentina, at a longitude of 69\fdg4 and a latitude of $-35\fdg2$. It has 1600 surface detectors (SDs) that cover an area of $3000 \unit{km}^2$, as well as four arrays of six atmospheric fluorescence telescopes. The \PAO\ recorded $\ncray = 27$ \uhecrs\ with reliable detected energies of $E_\obs \geq E_\uhemin = 5.7 \times 10^{19} \unit{eV}$ between 2004 January 1 and 2007 August 31 \citep{AugerCollab:2007b}.  These 27 events are shown in \fig{skymap}.  

The arrival directions of \uhecrs\ are measured with an accuracy of about $1 \unit{deg}$ \citep{AugerCollab:2008} by the \PAO, although there is an additional effective uncertainty in the progenitor directions as the \uhecrs\ are deflected by Galactic and inter-galactic magnetic fields.  The magnitude of this effect is somewhat uncertain, with estimates of the typical deflection angles ranging from $2 \unit{deg}$ (\eg\ \citealt{Dolag_etal:2005,MedinaTanco_etal:1998}) to $10 \unit{deg}$ (\eg\ \citealt{Sigl_etal:2004}) for $E_\uhemin \simeq 10^{20} \unit{eV}$ \uhecrs.  Despite lack of knowledge about the magnetic field strengths, the combined effect of both the deflection and the errors in the directional reconstruction is to ensure that the observed arrival direction, $\skyhat_\obs$, and the unit vector to the progenitor, $\skyhat_\source$, are separated by, typically, a smearing angle of a few degrees.  We model this process by defining the conditional probability distribution of observed arrival directions of \uhecrs\ from a source at $\skyhat_\source$ as a two-dimensional Gaussian on the sphere,
\[
\prob(\skyhat_\obs | \skyhat_\source) 
  = \frac{1}{2 \pi \sigma^2 (1 - e^{-2/\sigma^2})} 
    \exp\left(- \frac{1 - \skyhat_\obs \cdot \skyhat_\source}{\sigma^2} \right).
\]
\begin{equation}
\label{eq:psf}
\end{equation}
We assume a fiducial smearing angle of $\sigma = 3 \unit{deg}$ unless otherwise stated, but also calculate results using $\sigma = 6 \unit{deg}$ and $\sigma = 10 \unit{deg}$ for comparison purposes.

Over the 3.6 years that the 27 \uhecrs\ were detected, the effective area of the \PAO\ increased steadily, but the evolution was sufficiently gradual that the exposure per unit solid angle, $\diff \exposure / \diff \Omega$ (which has units of area $\times$ time) is a function of declination only.  The angular dependence of the \PAO\ exposure can be approximated by assuming that the instantaneous exposure is constant within $60 \unit{deg}$ of the zenith and zero otherwise.  (The detailed angular dependence is dominated by the cross-sectional area of the SD array, and there are smaller corrections due to the various \PAO\ data cuts, but these secondary effects are ignored here.)  Integrating the instantaneous exposure over time to account for the Earth's rotation (\cf\ \citealt{Fodor_Katz:2001}) yields the declination-dependent exposure $\exposure(\skyhat)$ shown in the left panel of \fig{skymap}. The total exposure, $\exposure_{\rmn{tot}} = \int (\diff \exposure / \diff \Omega) \; \diff \Omega$, for the \PAO\ observations considered here is $9000 \unit{yr} \unit{km}^2 \unit{sr}$ \citep{AugerCollab:2007a}.

\subsection{Local \agns}
\label{section:agn}

We follow \cite{AugerCollab:2007b} in considering only \agns\ in the 12th edition of the \cite{VeronCetty:2006} catalogue as possible sources for the \PAO\ \uhecrs.  The distance to each source, $D$, is calculated from the quoted absolute and apparent magnitudes in the \VCV\ catalogue, and \agns\ without absolute magnitudes are omitted.  The full \VCV\ catalogue contains $108\,014$ \agns, but only $\nsrc = 921$ have $z_\obs \leq 0.03$ and are hence plausible \uhecr\ progenitors inside the \gzk\ horizon of about $100 \unit{Mpc}$.

The \VCV\ catalogue is heterogeneous, having been compiled from a variety of \agn\ and quasar surveys and, as such, it is not ideal for statistical studies.  It is, however, expected to be close to complete for the local \agns\ of interest here, except close to the Galactic plane. Moreover, as emphasised in \sect{intro}, the \VCV\ sample was chosen specifically to facillitate comparison with the results of \cite{AugerCollab:2007b}.  


\section{Statistical method}
\label{section:method}

Do the observed arrival directions of the 27 \PAO\ \uhecrs\ provide evidence that at least some of them were emitted by known nearby \agns? We answer this question by using a two-component parametric model characterised by the rate at which \uhecrs\ are emitted by each \agn, $\ratesrc$, and the rate at which an isotropic background of \uhecrs\ arrive at Earth, $\ratebkg$\footnote{The two rates have different units: $\ratesrc$ is the average number of \uhecrs\ emitted per unit time by an \agn, and is given in units of $\unit{s}^{-1}$; $\ratebkg$ is the average number of background \uhecrs\ arriving at Earth per unit time, per unit area, per unit solid angle and is given in units of $\unit{s}^{-1} \unit{m}^{-2} \unit{sr}^{-1}$.}.  If none of the \uhecrs\ come from the candidate \agns\ then the data should be consistent with $\ratesrc = 0$; conversely, if all the \uhecrs\ come from the \agns\ in the catalogue, then the data should be consistent with $\ratebkg = 0$.  

The full constraints on $\ratesrc$ and $\ratebkg$ implied by the \PAO\ data are summarised in their joint posterior probability distribution, given by
\begin{equation}
\label{eq:post} 
\prob(\ratesrc, \ratebkg|\data) \nonumber \\
\end{equation}
\[
\mbox{}
  = \frac{ \prob(\ratesrc,\ratebkg) \, \prob(\data|\ratesrc, \ratebkg) }
    { 
      \int_{-\infty}^\infty \int_{-\infty}^\infty 
      \prob(\ratesrc, \ratebkg) \,
      \prob(\data|\ratesrc, \ratebkg)
     \, \diff \ratesrc \, \diff \ratebkg
  },
\]
where $\prob(\ratesrc,\ratebkg)$ is the prior distribution that encodes any external constraints on the rates, $\prob(\data|\ratesrc, \ratebkg)$ is the likelihood of obtaining the measured data given particular values for $\ratesrc$ and $\ratebkg$, and the integral in the denominator is the evidence. As we are not making any comparisons to other models, the only role the evidence plays here is to ensure the posterior is correctly normalised; hence it can be ignored when investigating the shape of the posterior. We adopt a uniform prior over $\ratebkg\ge0$ and $\ratesrc\ge0$, which plausibly encodes our ignorance of these parameters and also includes the a priori possible value of zero for both rates (unlike the Jefferys prior, uniform in the logarithm of the rates). Hence the following posterior plots also show the likelihood -- and, therefore, the constraining power of the \PAO\ data -- directly. Applying these simplifications, 
\eq{post} reduces to
\begin{equation}\label{eq:postfinal}
\prob(\ratesrc, \ratebkg|\data)
  \propto
  \Theta(\ratesrc) \Theta(\ratebkg)
   \prob(\data|\ratesrc, \ratebkg),
\end{equation}
where $\Theta(x)$ is the Heaviside step function.

A self-consistent statistical treatment requires the use of intrinsic source and background rates, although these parameters are not particularly intuitive themselves in the absence of a physical model for the production of \uhecrs. From those rates, however, we can calculate the expected number of source and background events in any sample, as well as the fraction of \uhecrs\ that have come from \agns, $\fagn$. The constraints on the expected \uhecr\ numbers are simply proportional to those on the relevant rates; $\fagn$ is given by the ratio of the expected number of \agn\ \uhecrs\ to the expected total number.  
It is crucial that we begin by parameterising our problem with the fundamental physical quantities, the rates $\ratesrc$ and $\ratebkg$, rather than the $\fagn$ (as done in \citealt{AugerCollab:2010b}); the latter is defined by the physical rates along with the energy range and observing footprint of a particular dataset, and hence the posterior distribution of $\fagn$ is derived from the posterior distribution of the rates, given in Eq.(~\ref{eq:postfinal}). 
%
%
%
Moreover, in small samples in which the total arrival rate of \uhecrs\ has a significant Poisson uncertainty, the only way to consistently account for the (independent) fluctuations in the source and background \uhecrs\ is to parameterise their rates explicitly.


\subsection{The likelihood}
\label{section:likelihood}

The likelihood is the probability of obtaining the observed data under the assumption of a particular model.  Here, the data take the form of the measured arrival directions, $\{\skyhat_\cray\}$, of the $\ncray$ \uhecrs\ (along with the value of $\ncray$ itself).  It is also possible to use the measured arrival energies of the \uhecrs, a possibility which is investigated in \cite{Mortlock_etal:2011c} but is not explored here. To evaluate the likelihood, we employ a `counts in cells' approach, dividing the sky into $\npx = 180 \times 360 = 64800$ pixels distributed uniformly in right ascension and declination.  The data are hence recast as the set of \uhecr\ counts in each pixel, $\databins$.  In the limit of infinitely small pixels this is mathematically equivalent to the likelihood written directly in terms of the arrival directions \citep{Mortlock_etal:2011c}, but is more straightforward to analyse and simulate.

The full likelihood of the data is a product of the independent Poisson likelihoods in each pixel, and is hence given by 
\[
\prob(\databins | \ratesrc, \ratebkg) 
\]
\begin{equation}  
\label{eq:likelihood} 
\mbox{}
  = \prod_{\px = 1}^{\npx} 
    \frac{(\ncrpxexpbkg + \ncrpxexpsrc)^{\ncrpx} 
    \exp[-(\ncrpxexpbkg + \ncrpxexpsrc)]}{\ncrpx!},
\end{equation}
where $\ncrpxexpbkg$ and $\ncrpxexpsrc$ are the expected number of background and source \uhecrs\ in pixel $\px$, respectively. In the limit of small pixels, the denominator in Eq.~(\ref{eq:likelihood}) can be ignored as $\ncrpx ! = 1$ if there is never more than one \uhecr\ in a pixel. 

The expected number of background \uhecrs\ in pixel $\px$ is 
\begin{equation} 
\label{eq:ncrpxexpbkg}
\ncrpxexpbkg = \ratebkg 
  \int_\px \frac{\diff \exposure}{\diff \Omega} \, \diff \Omega_\obs ,
\end{equation}
where the integral is over the $\px$'th pixel and $\diff \exposure / \diff \Omega$ is the \PAO\ exposure per unit solid angle (see \sect{auger}).  The expected number of \uhecrs\ from known sources in pixel $\px$ is 
\[
\ncrpxexpsrc 
\]
\begin{equation}
\label{eq:ncrpxexpsrc}
\mbox{} = 
   \sum_{s = 1}^{\nsrc}
    \frac{\diff N_\arr(E_\obs \geq E_{\uhemin}, D_\src)}{\diff t \, \diff A} 
  \int_\px \frac{\diff \exposure}{\diff \Omega} \,
    \prob(\skyhat_\obs | \skyhat_\src)
    \, \diff \Omega_\obs,
\end{equation}
where the sum is over the \agn\ sources, $\prob(\skyhat_\obs | \skyhat_\src)$ is the smearing probability (Eq.~\ref{eq:psf}), and ${\diff N_\arr(E_\obs \geq E_{\uhemin}, D)}/(\diff t \, \diff A)$ is the rate (\ie\ number per area and time) of \uhecrs\ from a source at distance $D$ arriving at Earth above the cut-off energy, $E_{\uhemin}$. This rate is therefore proportional to the source rate, $\ratesrc$, but further depends on both the shape of the \agn\ \cosmicray\ injection spectrum and also the distance-dependence of the \gzk\ energy losses, and so requires an explicit \uhecr\ model (see Eq.~\ref{eq:dNobs}).  

The positional dependence of $\ncrpxexpbkg$ and $\ncrpxexpsrc$ are both shown in \fig{skymap}.  The right panel is a combination of both the \PAO\ exposure and the local distribution of \agns, although comparing the left and the right panel it is clear that the latter dominates.  In particular, by far the strongest source is Centaurus A (with $l = 309\fdg5$ and $b = 19\fdg4$), which has previously been suggested as the dominant source of \uhecrs\ (\eg\ \citealt{AugerCollab:2007b}).  

It is possible within the Bayesian approach to assess whether any single \uhecr\ came from a particular source, and the full formalism for doing so is presented in \cite{Mortlock_etal:2011c}. However, a useful estimate of the probability that a \uhecr, with measured arrival direction in pixel $\px$, has come from one of the sources under consideration is 
\begin{equation}
\label{eq:psrc}
\psrc = 
\prob({\rmn{from \,\, source}} | p, \ratesrc, \ratebkg) 
  = \frac{\ncrpxexpsrc}{\ncrpxexpsrc + \ncrpxexpbkg},
\end{equation}
given values for the two rates.  As the rates inferred from a sample of even just 27 \uhecrs\ are not sensitive to any one event, it is reasonable to evaluate $\psrc$ using the best-fit values of $\ratebkg$ and $\ratesrc$ to assess the likely origin of each \uhecr\ in turn, and this is done for the \PAO\ data in \sect{results}.


\subsection{\uhecr\ model}
\label{section:model}

We adopt a simple model for \uhecr\ generation in which all \agns\ emit \uhecrs\ at the same overall rate and with a power-law energy flux of $J \propto E^{-\gamma}$.  This implies a differential emission rate of the form $\diff N_{\emit}/\diff E\propto E^{-\gamma-1}$.  The spectrum is normalised such that the total emission rate of \uhecrs\ with energy greater than $E$ is simply 
\begin{equation}\label{eq:dndedt_emit}
\frac{\diff N_\emit (> E)}{\diff t}
  = \ratesrc \left( \frac{E}{E_\uhemin} \right)^{- \gamma} ,
\end{equation}
where $E_\uhemin = 5.7 \times 10^{19} \unit{eV}$ is the minimum \uhecr\ energy and $\ratesrc$ is the rate at which each source emits \uhecrs\ (as above).  The \uhecr\ luminosity of each \agn\ is hence $L_{\rmn{src}} = \gamma / (\gamma - 1) \, \ratesrc E_\uhemin$.  We take $\ratesrc$ to be the same for all \agns, although it is plausible that the \uhecr\ emission rate scales with an \agn's hard X-ray luminosity (\eg\ \citealt{Protheroe_Szabo:1992}).  We also fix the logarithmic slope at $\gamma = 3.6$ \citep{AugerCollab:2010}.  Deviations from these fiducial models will be explored further in \cite{Mortlock_etal:2011c}. 

The dominant energy loss mechanism of \uhecrs\ is the interaction with the CMB photons through the GZK effect.  Although clearly a stochastic process, its most important feature is the exponential decrease in a \uhecr's energy with distance.  This can be captured by adopting a continuous loss approximation (\cf\ \citealt{Achterberg_etal:1999}) in which a \uhecr's arrival energy is given by
\begin{equation}
\label{eq:energy}
E_\arr = \max\left[ E_{\gzk}, E_\emit(1-f_{\gzk})^{D/L_{\gzk}} \right],
\end{equation}
where $D$ is the distance to the source, $f_{\gzk} = 0.2$ is the average fractional energy loss per \gzk\ interaction, and $L_{\gzk} = 4 \unit{Mpc}$ is the \gzk\ mean free path (\eg\ \citealt{Rachen_Biermann:1993}). It is also assumed that there are no further energy losses once a \cosmicray\ reaches $E_{\gzk}$, although this is unimportant for \uhecrs\ with $E_\uhemin > E_{\gzk}$ (such as those in the \PAO\ sample analysed here).

Combining \eqs{dndedt_emit} and (\ref{eq:energy}) with the distance to the \agn\ then gives the rate per unit area of \crs\ arriving at Earth with energy $E_\arr \geq E_\uhemin$ from each \agn\ as
\begin{equation}\label{eq:dNobs}
\frac{\diff N_\arr(E_\arr \geq E_{\uhemin}, D)}{\diff t \, \diff A}
 = \ratesrc \frac{(1-f_\gzk)^{\gamma D/L_{\gzk}}}{4\pi D^2}\;.
\end{equation}
This can finally be used in \eq{ncrpxexpsrc} to calculate the expected number of \crs\ in each pixel, and therefore the likelihood (given in Eq.~\ref{eq:likelihood}).


\subsection{Simulations}
\label{section:simulations}

It is useful to test the constraining power of a small number of \uhecrs\ by generating mock \PAO\ samples with known progenitor properties.  We created simulations of the two extreme cases: an \agn-only sample in which all the \uhecrs\ were sourced from the nearby \VCV\ \agns\ and propagated using the simple \gzk\ model described in \sect{model}; and an all-background sample in which the arrival directions are random over the whole sky. In both cases the incident \uhecrs\ were subject to the \PAO's measurement errors and declination-dependent exposure. Both samples were constrained to have exactly 27 events so as to provide parameter constraints that can be compared directly with those from the real \PAO\ sample\footnote{It would be inconsistent to draw $\ncray$ from a Poisson distribution with mean of 27, as the observed number of \uhecrs\ is already the result of a Poisson draw from the (unknown) mean number expected. One of the more convenient aspects of Bayesian parameter estimation is that it is possible to obtain error estimates without the need for an ensemble of realisations.}.

\begin{figure}
\includegraphics[width=\figwidth]{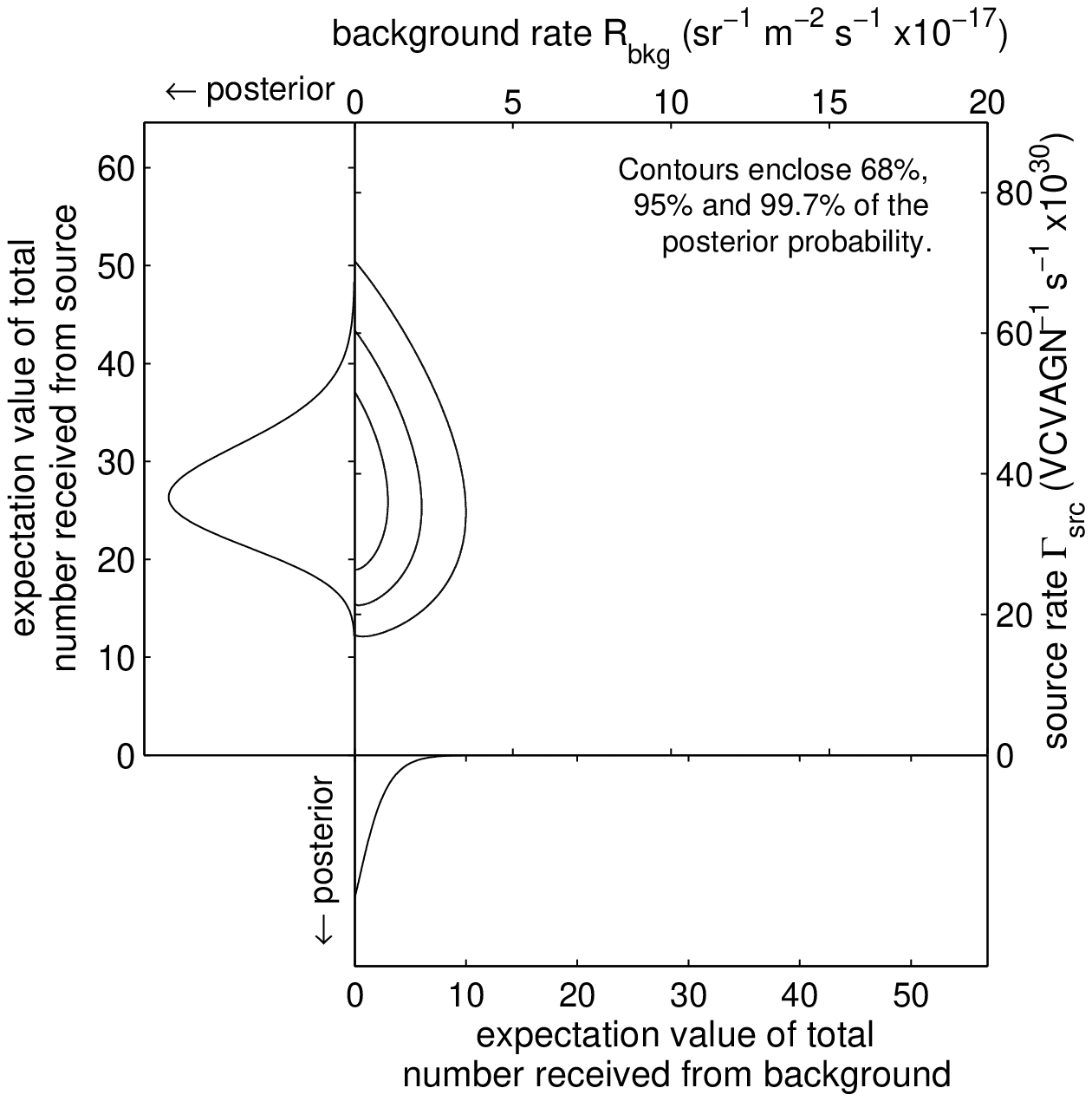}
\caption{The posterior probability of the \uhecr\ rate from \VCV\ \agns, $\ratesrc$, and the uniform background rate, $\ratebkg$, implied from a simulated sample of 27 \uhecrs, all of which were emitted by \VCV\ \agns.  The contours enclose 68\percent, 95\percent\ and 99.7\percent\ of the posterior probability, and the line plots show the marginalised probability for each rate.}
\label{figure:postpdf_simdata_agn}
\end{figure}

The results of the \agn-only simulation are shown in \fig{postpdf_simdata_agn}.  As expected, the constraints on $\ratesrc$ match the naive Poisson expectation; more interesting is the rejection of the possibility that more than a few of the \PAO\ \uhecrs\ are not from the \VCV\ \agns.  The constraints on the \agn\ fraction (see \fig{agnfraction}) from such a data-set would be $\fagn = 1.00_{-0.07}^{+0.00}$, where the quoted value is the maximum of the posterior, and all limits given in this paper enclose the most probable 68\percent\ of the posterior. This strong result implies that, if \agns\ source all \uhecrs, even a sample of 27 events would be sufficient to confirm this hypothesis if a complete catalogue of the progenitors was available.  

\begin{figure}
\includegraphics[width=\figwidth]{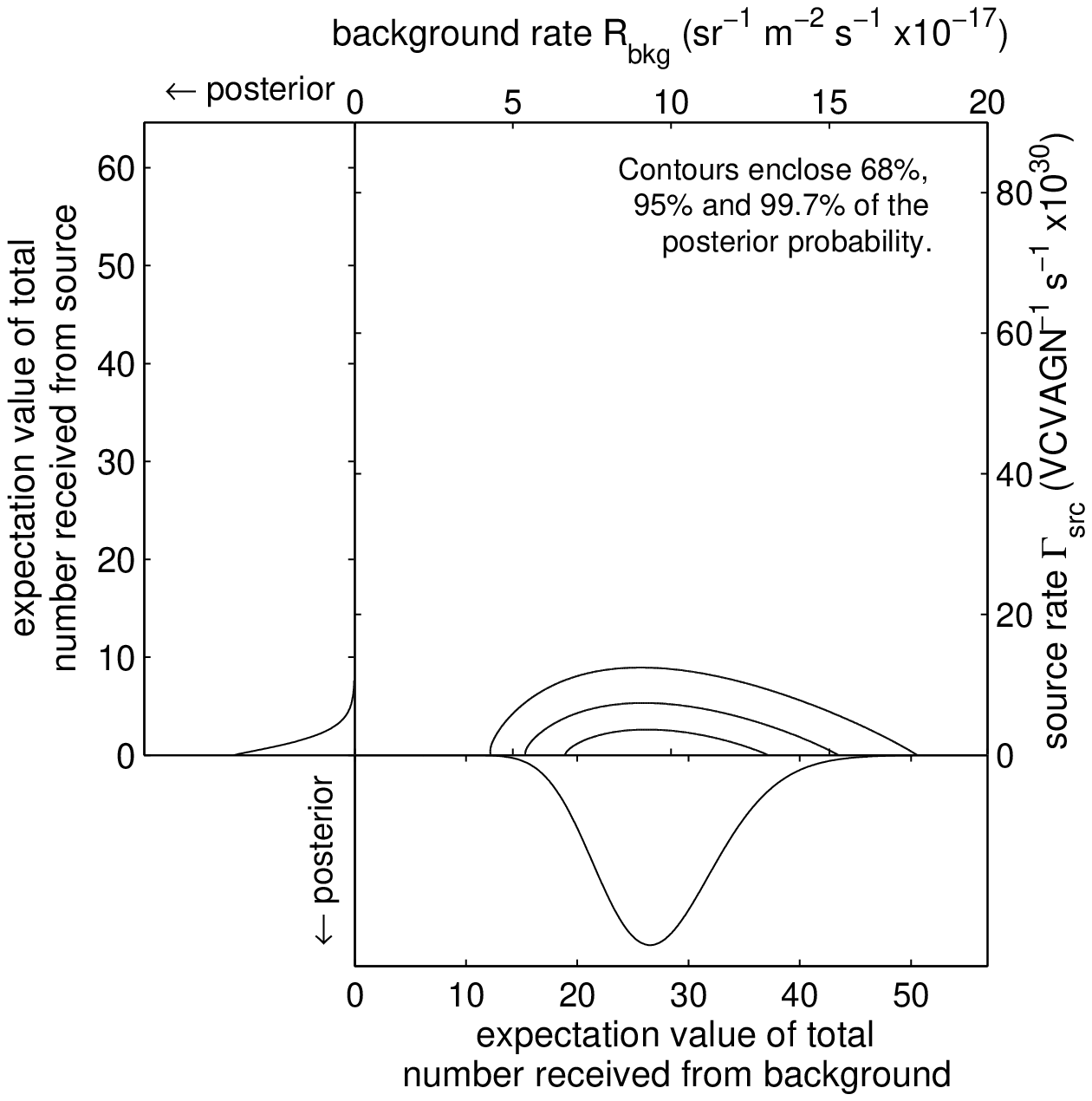}
\caption{Same as \fig{postpdf_simdata_agn}, but for a simulated sample of 27 isotropically distributed \uhecrs.}
\label{figure:postpdf_simdata_bkg}
\end{figure}

The results of the background-only simulation are shown in \fig{postpdf_simdata_bkg}.  Again, the constraints on $\ratebkg$ match the Poisson expectation.  The resultant constraints on the \agn\ fraction would be $\fagn = 0.00_{-0.00}^{+0.07}$ (see \fig{agnfraction}). However it is also important to note that some pixels (very far away from any AGN) have negligible contribution from the \VCV\ \agns, and because some of the \uhecrs\ in this sample fell in those pixels, there is a strong upper bound on $\fagn$ that is significantly lower than unity.

The fact that the posteriors from the \agn -only and the background-only simulations are almost completely disjoint implies that even a sample of just 27 \uhecrs\ might be sufficient to provide a definitive answer as to their origin. Given the observed distribution of the \PAO\ \uhecrs, the parameter constraints from the real data should lie between the two extremes shown in \figs{postpdf_simdata_agn} and \ref{figure:postpdf_simdata_bkg}.


\section{Results}
\label{section:results}

\begin{figure}
\includegraphics[width=\figwidth]{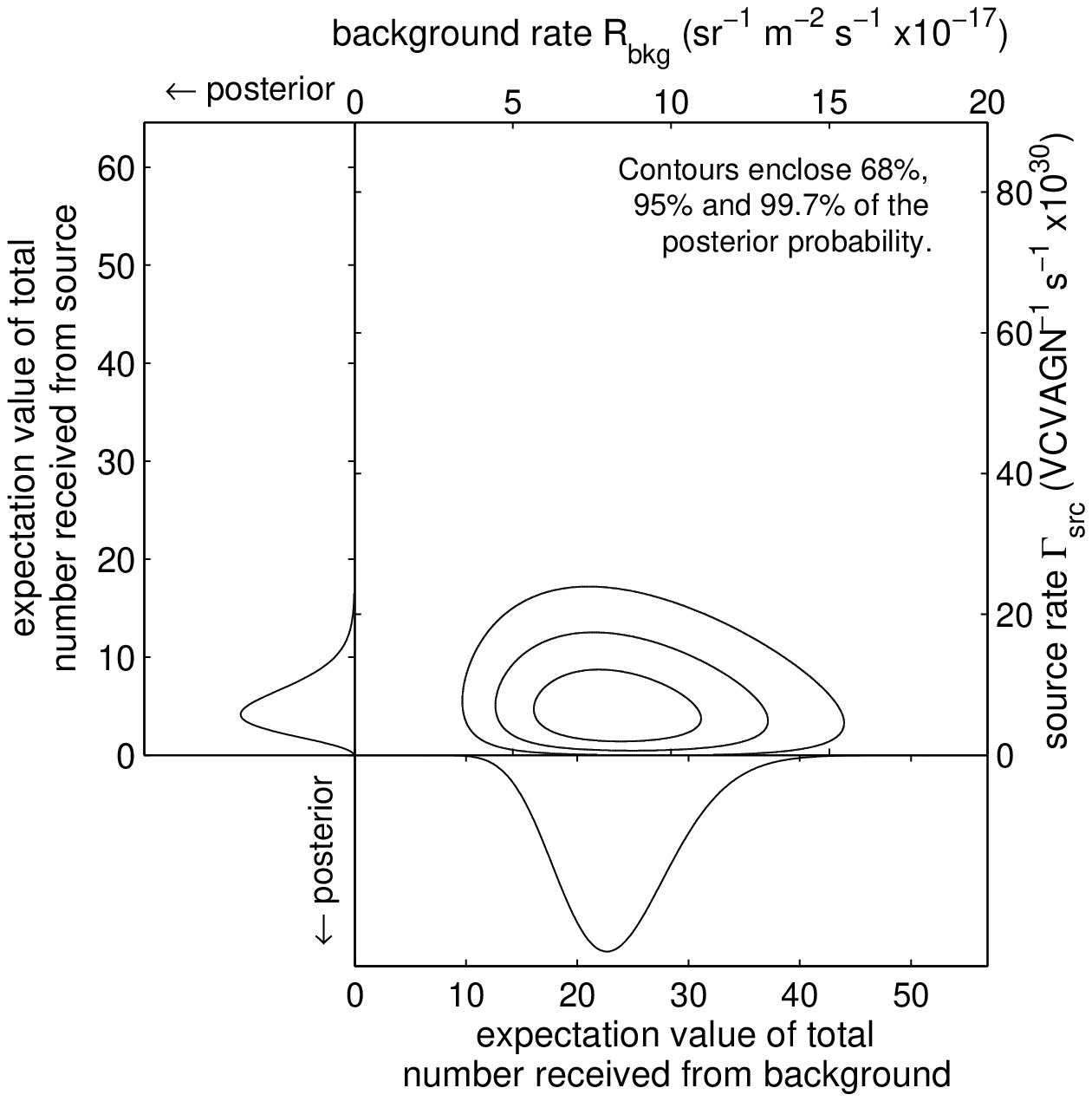}
\caption{Same as \fig{postpdf_simdata_agn}, but for all 27 \PAO\ \uhecrs.}
\label{figure:postpdf_realdata}
\end{figure}

\begin{figure}
\includegraphics[width=\figwidth]{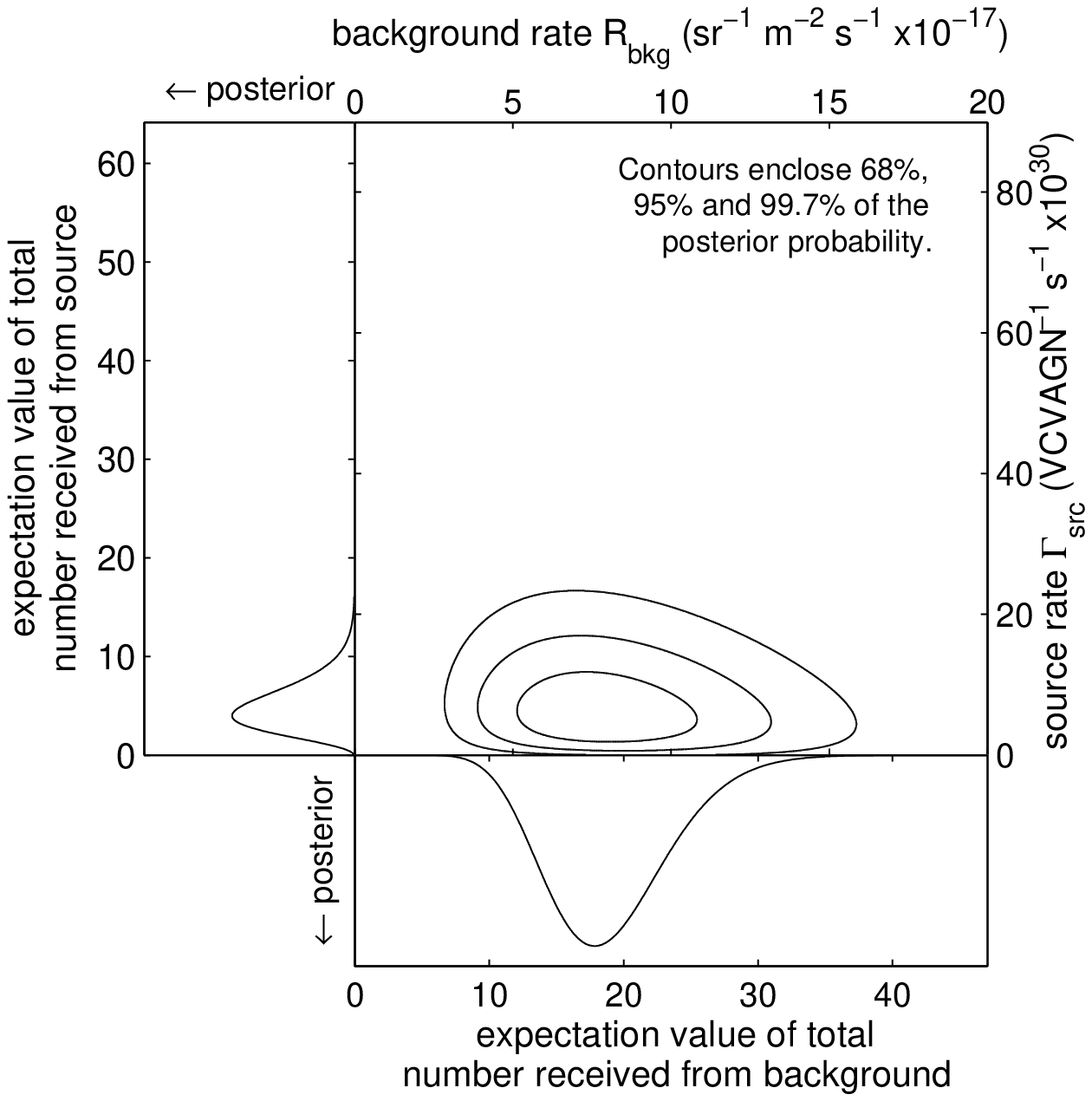}
\caption{Same as \fig{postpdf_simdata_agn}, but for the 22 \PAO\ \uhecrs\ which with arrival directions at least $10 \unit{deg}$ from the Galactic plane.}
\label{figure:postpdf_cutdata}
\end{figure}

\begin{figure}
\includegraphics[width=\figwidth]{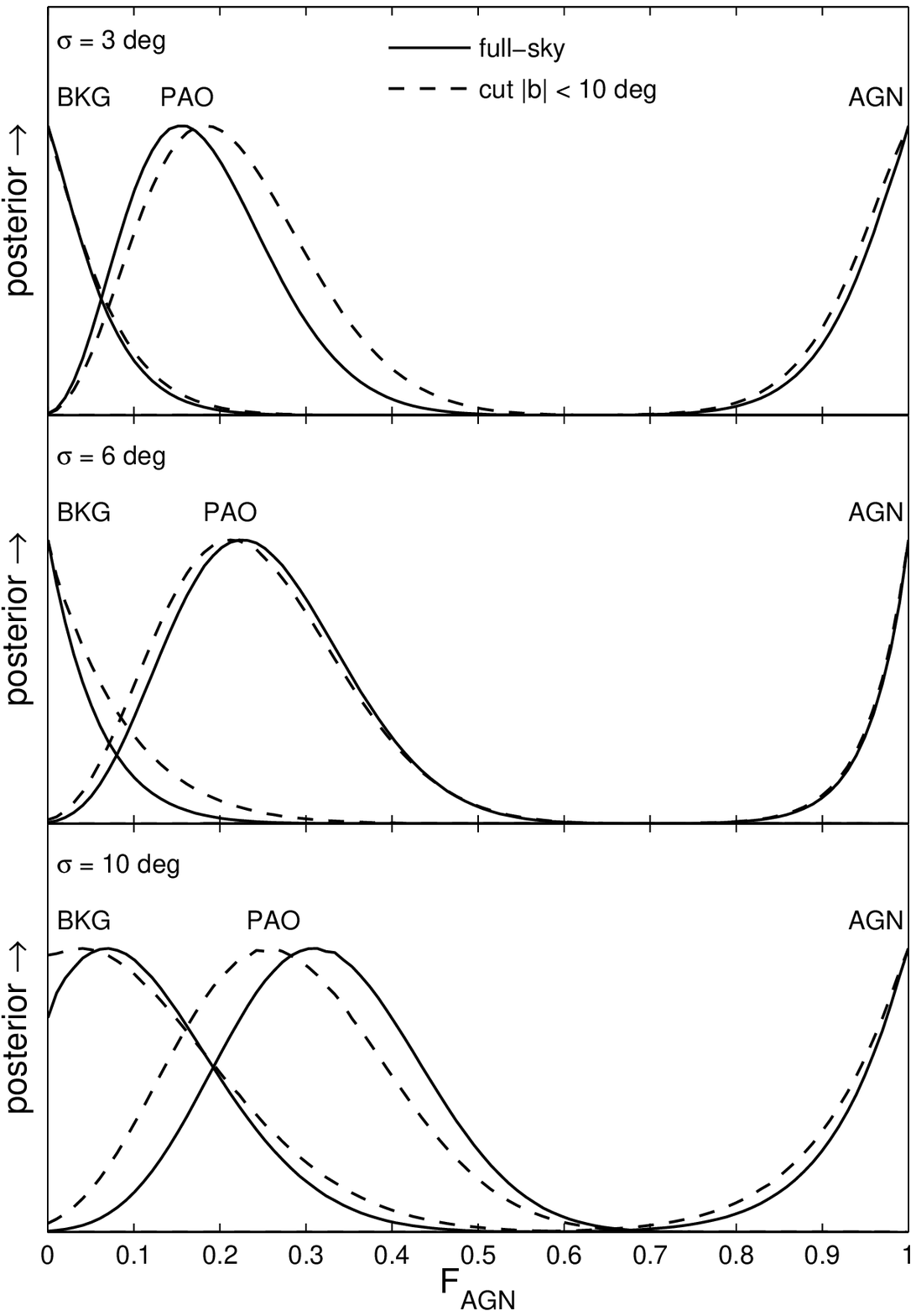}
\caption{Posterior distributions of the fraction of observed \uhecrs\ that are from the population of \VCV\ \agns, $\fagn$, shown for simulated samples (both isotropic and \agn-only) and for the real \PAO\ data. Curves for both the full sample of 27 \uhecrs\ and the cut sample of 22 \uhecrs\ (with arrival directions at least $10 \unit{deg}$ from the Galactic plane) are shown in the same panel. Each panel represents a different smearing angle.}
\label{figure:agnfraction}
\end{figure}

The posterior probability distribution in $\ratesrc$ and $\ratebkg$ given the \PAO\ \uhecr\ sample is shown in \fig{postpdf_realdata}.  The constraints in this figure represent our main result and it is useful to discuss some of its features in more detail.  As expected, the posterior is intermediate between the extreme cases shown in \figs{postpdf_simdata_agn} and \ref{figure:postpdf_simdata_bkg}.  We find the marginalised rates to be $\ratesrc = (5.8^{+4.0}_{-2.9})\times10^{30}\unit{s}^{-1}$ (equivalent to \uhecr\ source luminosity of $L_{\rmn{src}} = 7.4^{+5.1}_{-3.7} \times 10^{31} \unit{W}$) and $\ratebkg = (8.0^{+1.9}_{-1.6})\times10^{-17}\unit{sr}^{-1}\unit{m}^{-2}\unit{s}^{-1}$. We also calculate the posterior distribution of the fraction of the \PAO\ \uhecrs\ that come from \VCV\ \agns, shown in \fig{agnfraction}. The most probable value is $\fagn = 0.15$, and the constraints can be summarised by the interval $\fagn=0.15^{+0.10}_{-0.07}$ (where these limits enclose the most likely $68\percent$ of the posterior probability).  

As most extragalactic catalogues are incomplete close to the Galactic plane we also repeated the above analysis on a reduced data-set from which the region with Galactic latitudes of $|b| \leq 10 \unit{deg}$ had been removed (see \fig{postpdf_cutdata}).  The \PAO\ exposure in the retained regions is $7480 \unit{yr} \unit{km}^2 \unit{sr}$ and the number of \uhecrs\ included was reduced from 27 to 22.  The lower numbers resulted in slightly broader constraints on $\fagn$, as can be seen from \fig{agnfraction}. From this cut data, we find $\ratesrc = (5.6^{+3.9}_{-2.8})\times10^{30}\unit{s}^{-1}$, $\ratebkg = (7.6^{+2.0}_{-1.7})\times10^{-17}\unit{sr}^{-1}\unit{m}^{-2}\unit{s}^{-1}$ and $\fagn=0.18^{+0.11}_{-0.09}$.

The analysis was also repeated using larger mean smearing angles of $\sigma = 6 \unit{deg}$ and $\sigma = 10 \unit{deg}$.  The limits on the \agn\ fraction in these models are $\fagn = 0.22^{+0.12}_{-0.09}$ ($\sigma = 6 \unit{deg}$) and $\fagn = 0.31^{+0.12}_{-0.12}$ ($\sigma = 10 \unit{deg}$).  In both cases the most probable value of $\fagn$ is higher than in the fiducial model, although a broader range of $\fagn$ values is compatible with the data as well.  It is natural that a higher \agn\ fraction be compatible with the data given larger values of $\sigma$, as a greater fraction of the sky is within $\sigma$ of at least one source, and this effect has been seen by \eg\ \cite{Kim_Kim:2011} and \cite{AugerCollab:2010}.  In particular, \cite{Kim_Kim:2011} report the fraction of observed \uhecrs\ that originate from \agns\ to be $0.45$ for a smearing angle of $6\unit{deg}$. However the best-fit value of $\fagn$ increases less strongly with $\sigma$ in the Bayesian formalism we present than found by using other methods; the inherent self-consistency of the Bayesian approach ensures the correct balance is struck between the compatibility of this more forgiving model and the lack of predictivity.  

There is strong evidence of a \uhecr\ signal from the known \VCV\ \agns, which manifests in the result that $\fagn = 0.15$ is 200 times more probable than $\fagn = 0.00$, but not all the \PAO\ \uhecrs\ can be explained this way.  These results could also be cast in terms of model comparison if only the background-only (\ie\ $\fagn = 0$) or the \agn-only (\ie\ $\fagn = 1$) possibilities were considered.  The former case, which is the null hypothesis rejected by \cite{AugerCollab:2007b}, is actually reasonably consistent with the data, whereas the hypothesis that all the \PAO\ \uhecrs\ come from \VCV\ \agns\ is completely ruled out because there are several events with no plausible \agn\ progenitor in the \VCV\ catalogue.  The probability that each of the 27 \uhecrs\ came from one of the \VCV\ \agns\ was calculated explicitly according to \eq{psrc} by adopting the best-fit values for $\ratebkg$ and $\ratesrc$ given above; these probabilities are given in \tabl{crs}.  For $\sigma=3 \unit{deg}$, only 9 events have $\psrc \ga 0.1$, all of which were identified as being within $3.2 \unit{deg}$ of an \agn\ with $\redshift \leq 0.017$ by \cite{AugerCollab:2007b}.  However the other 11 events which \cite{AugerCollab:2007b} identified as \agn\ correlated have very low values of $\psrc$, in most cases because the angular correlation is with an \agn\ that is close to their maximum redshift and so has a significantly reduced \uhecr\ flux at Earth.  Moreover, 14 of the \uhecrs\ have $\psrc < 0.001$, with no plausible \agn\ progenitor, at least within the \VCV\ catalogue. As also shown in \tabl{crs}, the results are similar, but less conclusive, for larger smearing angles. The \agn\ hypothesis cannot be ruled out for the low $\psrc$ events, however: these \uhecrs\ could have come from \agns\ that are not in the \VCV\ catalogue (and some could have come from \VCV\ \agns\ if deflected by more than a few degrees).

\begin{table}
\centering
\caption{The measured arrival directions of the 27
  \PAO\ \uhecrs\ listed in Abraham \etal\ (2007b) along with their
  assessment of of \agn\ correlation (PAO) and our values of the \agn\
  progenitor probability (which is rounded to zero if less than 0.0005) 
  for the three different smearing angles. The \crs\ marked with *1, *2 and *3 
  in the {\it b} column are those closest to Centaurus A, with angular 
  separations of 0.9, 2.3 and 5.8 deg respectively.}
\label{table:crs}
\begin{tabular}{rrcccc}
\hline
$l$ & $b\;\;\:\,$ & PAO & $P_{\rmn{src}}$ & $P_{\rmn{src}}$ & $P_{\rmn{src}}$ \\
$\rmn{deg}$ & $\rmn{deg}\;\;\:\,$ & corr.\ & $\!\sigma \!=\! 3 \unit{deg}\!$ & $\!\sigma \!= \!6 \unit{deg}\!$ & $\!\sigma \!= \!10 \unit{deg}\!$ \\
\hline
$ 15.4$ & $  8.4\;\;\:\,$ & no & 0.000 & 0.000 & 0.000\\
$-50.8$ & $ 27.6\;\;\:\,$ & yes & 0.559 & 0.761 & 0.681\\
$-49.6$ & $  1.7\;\;\:\,$ & yes & 0.000 & 0.134 & 0.387\\
$-27.7$ & $-17.0\;\;\:\,$ & yes & 0.099 & 0.067 & 0.033\\
$-34.4$ & $ 13.0\;\;\:\,$ & yes & 0.078 & 0.171 & 0.424\\
$-75.6$ & $-78.6\;\;\:\,$ & yes & 0.380 & 0.528 & 0.493\\
$ 58.8$ & $-42.4\;\;\:\,$ & yes & 0.000 & 0.000 & 0.000\\
$-52.8$ & $ 14.1^{*3}$ & yes & 0.870 & 0.836 & 0.711\\
$  4.2$ & $-54.9\;\;\:\,$ & yes & 0.000 & 0.004 & 0.008\\
$ 48.8$ & $-28.7\;\;\:\,$ & yes & 0.000 & 0.000 & 0.000\\
$-103.7$ & $-10.3\;\;\:\,$ & no & 0.000 & 0.000 & 0.001\\
$-165.9$ & $-46.9\;\;\:\,$ & yes & 0.000 & 0.003 & 0.010\\
$-27.6$ & $-16.5\;\;\:\,$ & yes & 0.099 & 0.067 & 0.033\\
$-52.3$ & $  7.3\;\;\:\,$ & no & 0.167 & 0.533 & 0.577\\
$ 88.8$ & $-47.1\;\;\:\,$ & yes & 0.000 & 0.000 & 0.002\\
$-170.6$ & $-45.7\;\;\:\,$ & yes & 0.000 & 0.006 & 0.011\\
$-51.2$ & $ 17.2^{*2}$ & yes & 0.952 & 0.873 & 0.735\\
$-57.2$ & $ 41.8\;\;\:\,$ & no & 0.005 & 0.123 & 0.294\\
$ 63.5$ & $-40.2\;\;\:\,$ & yes & 0.000 & 0.000 & 0.000\\
$-51.4$ & $ 19.2^{*1}$ & yes & 0.964 & 0.881 & 0.742\\
$-109.4$ & $ 23.8\;\;\:\,$ & yes & 0.000 & 0.000 & 0.002\\
$-163.8$ & $-54.4\;\;\:\,$ & yes & 0.001 & 0.006 & 0.020\\
$-41.7$ & $  5.9\;\;\:\,$ & no & 0.002 & 0.208 & 0.454\\
$ 12.1$ & $-49.0\;\;\:\,$ & yes & 0.000 & 0.001 & 0.003\\
$-21.8$ & $ 54.1\;\;\:\,$ & yes & 0.000 & 0.005 & 0.088\\
$-65.1$ & $ 34.5\;\;\:\,$ & no & 0.000 & 0.049 & 0.321\\
$-125.2$ & $ -7.7\;\;\:\,$ & no & 0.001 & 0.002 & 0.002\\
\hline
\end{tabular}

\end{table}

\section{Conclusions}
\label{section:conc}
 
We have performed a Bayesian analysis to test whether the first 27 \uhecrs\ with $E_\obs \geq 5.7 \times 10^{19} \unit{eV}$ detected by the \PAO\ (i.e. those observed before 2007 September) have come from the known local \agns\ in the \VCV\ catalogue.  The first main conclusion from this analysis is that at least some do -- or at least come from progenitors within a few degrees of the \VCV\ \agns.  The fraction of \uhecrs\ that come from the \VCV\ \agns\ is constrained to be $0.15_{-0.07}^{+0.10}$. Conversely, our second important result is that many of the \PAO\ \uhecrs\ have not come from \agns\ in the \VCV\ catalogue, either because of incompleteness (most obviously close to the Galactic plane) or because there is another source of \uhecrs, possibly in our own Galaxy.

Our results differ somewhat from those presented by \cite{AugerCollab:2007b} due to our more explicit modelling of background and source events as well as the different statistical methods used.  The starting point of their analysis is the null hypothesis that the \uhecrs\ have not come from local \agns; they find that this is rejected `at the 99\% level' given the number of the \uhecrs\ that are within $3 \unit{deg}$ of a \VCV\ \agn.  This result was as expected, which illustrates the potentially circular reasoning when the obvious simple null hypothesis does not match prior knowledge (\ie\ the expectation that some of the \uhecrs\ did, in fact, come from the \agns), although the strength of the correlation made it clear that there was at least some connection between the two populations.  But it is impossible to go beyond this limited statement due to the use of arbitrary cuts in the correlation analysis (both in angular radius and \agn\ redshift), the equal weighting of the nearest known \agn, Centaurus A, with the hundreds of \agns\ at distances of about $100 \unit{Mpc}$, and the equal value placed on any angular match out to about $3 \unit{deg}$, which dilutes whatever correlation signal is present (and also increases the chance that a non-\agn\ sourced \uhecr\ is assigned a spurious match).  The simulations of \agn -only and background-only \uhecr\ samples shown here demonstrate that even a sample of just 27 events is sufficient to decisively distinguish between these two extreme possibilities, but also that the apparent strong correlation inferred by \cite{AugerCollab:2007b} is in part due to the analysis method. By introducing a model of both the \agn-sourced \uhecrs\ and a uniform background, the Bayesian analysis can be thought of as giving the optimal weight to any potential \uhecr-\agn\ pairing, given our prior knowledge of the physics of \uhecr\ propagation and the measurement process. 

During the final preparation of this paper, the Pierre Auger Collaboration presented an extended analysis of an enlarged set of 69 \uhecrs\ (\citealt{AugerCollab:2010b}), also comparing the \uhecr\ arrival directions with other extragalactic catalogues. Aside from the correlation-based methods they had used previously, they also included a likelihood-based formalism that has some similarities to our method. The results of the two approaches are broadly similar (whilst differing from the earlier correlation-based analyses), primarily because they both include a physical model of \uhecr\ propagation. They hence go closer to the ideal of including all the available information (\ie\ not just the data but knowledge of the CR physics) and so produce more robust results.

In addition to applying the methods described here to this enlarged dataset, there are several extensions to our analysis 
that will allow more rigorous conclusions regarding the origins of these particles.  Most importantly, we can account for the energy of individual \crs\ in our likelihood, rather than just demanding they are above the $E_\uhemin = 5.7 \times 10^{19} \unit{eV}$ cut.  This, in turn, will make it more important to use a more realistic, stochastic calculation of the \gzk\ effect, and also the energy-dependent \cosmicray\ deflection due to magnetic fields.  It will similarly be more important to investigate the possibility that the \agn\ \uhecr\ emission rate scales with \agn\ luminosity; a corollary is that it may be possible to discriminate between different \agn\ emission models.

In future work, we will investigate whether \uhecr\ data could be used yet more efficiently by including lower-energy events.  This would obviously increase the numbers, although there is the potentially severe penalty of diluting the angular signal by including \uhecrs\ that have either been deflected by more than about $10 \unit{deg}$ or have come from the many \agns\ at distances of greater than about $100 \unit{Mpc}$.  To the degree that the \cosmicray\ propagation and deflection models are accurate, these trade-offs can be evaluated objectively, following the underlying principle of extracting as much information as possible from the \uhecr\ measurements \citep{Mortlock_etal:2011c}.

Another way to potentially make better use of \uhecr data would be to use a more homogeneous \agn\ sample than the \VCV\ catalogue. An obvious example is the catalogue of \agns\ from the {\em{Swift}} Burst Alert Telescope (BAT) survey \citep{Tueller_etal:2008}, which has nearly uniform selection criteria outside the Galactic plane. Both \cite{George_etal:2008} and the latest \PAO\ analysis from \cite{AugerCollab:2010b} compare \uhecr\ data to this catalogue. In particular, \cite{George_etal:2008} approach the analysis in a fashion similar to that of \cite{AugerCollab:2007b} and found correlation at the `98\% level'. \cite{Mortlock_etal:2011c} will apply the fully Bayesian methods described in this paper to the BAT \agns, as well as extending the approach in order to provide a more rigorous analysis. By combining optimal statistical methods with the ever-increasing \uhecr\ data-sets it should soon be possible to definitively determine the origins of \uhecrs.


\section*{Acknowledgments}

This research would not have been possible without the help of several members of the Pierre Auger Collaboration, particularly Johannes Knapp, Angela Olinto, Benjamin Rouill\'{e} d'Orfeuil and Subir Sarkar. Abraham Achterberg's help was also invaluable, in particular by providing an unpublished manuscript. Roberto Trotta provided valuable input during the early stages of this project.  This paper was improved thanks to a number of thoughtful comments by the anonymous referee.


\bibliographystyle{mn2e.bst}
\bibliography{references_mnras}


\bsp
\label{lastpage}
\end{document}